%% file: nuclearspin.tex
\documentclass[aps,pra,twocolumn]{revtex4-1}

\usepackage{amsmath,amsfonts,amssymb}
\usepackage{dcolumn}

\include{phys_jdf}

\newcommand{\Sp}{\ensuremath{\left< s_{p} \right>}}
\newcommand{\Sn}{\ensuremath{\left< s_{n} \right>}}
\newcommand{\LQCD}{\ensuremath{\Lambda_\textit{QCD}}}
\newcommand{\MeV}{\textrm{MeV}}
\newcommand{\vect}[1]{\ensuremath{\boldsymbol{#1}}}
\newcommand{\eref}[1]{(\ref{#1})}
\newcommand{\Eref}[1]{Eq.~(\ref{#1})}
\newcommand{\Tref}[1]{Table~\ref{#1}}
\newcommand{\Sec}[1]{Section~\ref{#1}}

\begin{document}

\title{Search for variation of fundamental constants and violations of fundamental symmetries using isotope comparisons}

\author{J. C. Berengut}
\author{V. V. Flambaum}
\author{E. M. Kava}
\affiliation{School of Physics, University of New South Wales, Sydney, NSW 2052, Australia}

\date{9 September 2011}


\begin{abstract}

Atomic microwave clocks based on hyperfine transitions, such as the caesium standard, tick with a frequency that is proportional to the magnetic moment of the nucleus. This magnetic moment varies strongly between isotopes of the same atom, while all atomic electron parameters remain the same. Therefore the comparison of two microwave clocks based on different isotopes of the same atom can be used to constrain variation of fundamental constants. In this paper we calculate the neutron and proton contributions to the nuclear magnetic moments, as well as their sensitivity to any potential quark mass variation, in a number of isotopes of experimental interest including $^{201,199}$Hg and $^{87,85}$Rb, where experiments are underway. We also include a brief treatment of the dependence of the hyperfine transitions to variation in nuclear radius, which in turn is proportional to any change in quark mass. Our calculations of expectation-values of proton and neutron spin in nuclei are also needed to interpret measurements of violations of fundamental symmetries.

\end{abstract}

\maketitle

\section{Introduction}

Studies investigating possible space-time variation of the fundamental constants of particle physics have been given additional motivation in recent years by indications of variation of the fine-structure constant, $\alpha = e^2/\hbar c$, in quasar absorption systems~\cite{webb99prl,murphy03mnras,webb10arxiv}. Such a variation could be confirmed using complementary astrophysical~\cite{berengut11prd} and laboratory~\cite{berengut10arxiv0} studies, but so far all laboratory studies have shown results consistent with zero variation (see, e.g. the reviews~\cite{flambaum09ijmpa,berengut11jpcs}).

Theoretical models that attempt to unify the forces of the Standard Model (electroweak and strong nuclear) with gravity consistently predict that a variation in $\alpha$ would be related to a variation in the quantum chromodynamics (QCD) scale and the Higgs field. Such variations could be measured as a change in the dimensionless ratio $m_q/\LQCD$, where $m_q$ is the light quark mass (formally the average of the up and down quark masses, $m_q = (m_u + m_d)/2$) and $\LQCD$ is the quantum chromodynamics constant (which shows the position of the pole in the running strong coupling constant). The variation of $m_q/\LQCD$ has been estimated to be as great as 35 times that of $\alpha$, and consequently may be more suitable for investigation~(see, e.g.~\cite{marciano84prl,langacker02plb,calmet02epjc,wetterich03jcap,dent03npb}).

Measuring a change in $m_q/\LQCD$ directly is not possible, so some effort must be made to calculate the relationship between variation in quark mass and that of an observable quantity. The dependence of proton and neutron magnetic moments on $m_q/\LQCD$ has previously been calculated~\cite{flambaum04prd,cloet08fbs}. These calculations in turn can be used to obtain the sensitivity of nuclear magnetic moments, $\mu$, to variations in $m_q/\LQCD$~\cite{flambaum06prc}. Nuclear magnetic moments can be probed by the hyperfine structure of atomic clocks, which form the reference transition in clocks such as the caesium frequency standard.

By comparing the hyperfine frequencies of isotopes of the same atom, a stringent limit can be placed on variations in $m_q/\LQCD$. The comparison is insensitive to $\alpha$-variation since this dependence comes only from the electronic structure, which is common to both. For the same reason, difference measurements between isotopes of the same ion have strongly reduced sensitivity to pressure shifts, second-order Zeeman shifts, and second-order Doppler shifts~\cite{burt09fsm}. Experiments currently underway include comparison of $^{201}$Hg and $^{199}$Hg clocks~\cite{burt09pra,burt09fsm}, and $^{87}$Rb versus $^{85}$Rb clocks~\cite{wei11private}; for limits from other microwave clock experiments see the review~\cite{lea08epjst}. Other candidate isotopes are also considered in this paper. In this work we assume that $\LQCD$ is constant, calculating the dependence on the small parameter $m_q$. This is not an approximation: rather it means that we measure $m_q$ in units of $\LQCD$. Therefore $\delta m_q/m_q$ should be understood as a relative change in $m_q/\LQCD$.


In this work we have two aims. First, we find expectation values of proton and neutron spin for isotopes of nuclei of experimental interest. These are needed in order to calculate nuclear electric dipole moments (EDMs)
\begin{equation}
d_N \frac{I_z}{I} = d_p\frac{\Sp_z}{s_p} + d_n \frac{\Sn_z}{s_n}
\end{equation}
due to proton and neutron EDMs ($d_p$ and $d_n$, respectively). Here $I$ is the nuclear spin and $s_p=s_n=1/2$. Similarly we may calculate constants of the parity non-conserving (PNC) nuclear-spin-dependent interaction between electrons and the nucleus~(see, e.g.~\cite{ginges04prep,khriplovich91book}) using
\begin{equation}
\varkappa = \varkappa_p \frac{\Sp}{s_p} + \varkappa_n\frac{\Sn}{s_n}
\end{equation}
where $\left< s \right> = \left< s \right>_z$ for the maximal projection $I_z=I$; we omit the subscript $z$ for brevity. Values of \Sp\ and \Sn\ are presented for many nuclei of experimental interest in Tables~\ref{tab:spin_proton} and \ref{tab:spin_neutron}. In these tables the ``BI'' column represents our preferred value, while the difference between BI and BII gives an indication of the uncertainty in the calculation.

Our second aim is to use these expectation values to calculate the sensitivity of the nuclear magnetic moment to variation of quark mass, denoted $\kappa$. We then improve our estimates of $\kappa$ by including the variation in proton and neutron-spin expectation values induced by quark mass variation, as well as nuclear radius effects. Results for nuclei of experimental interest are presented in \Tref{tab:kappa}. We identify potential candidate nuclei for experimental study where there is high sensitivity to quark-mass variation and good stability of calculations across the various theoretical models.

\section{Nuclear magnetic moment}

The first step in our calculations is to disentangle the proton and neutron contributions to the nuclear magnetic moment (in units of the nuclear magneton $\mu_{N} = e\hbar/2m_{N}$):
\begin{equation}
\label{eq:mu_general}
\mu = g_{p} \Sp + g_{n} \Sn + \left< l_{p} \right> \,,
\end{equation}
where \Sp\ and \Sn\ are the expectation values of the total proton and neutron spins, while $\left< l_{p} \right>$ is the orbital angular momentum of an unpaired valence proton (if there is one). In this paper we treat the case where the nucleus has either one valence neutron or one valence proton: because of nuclear pairing even-even nuclei have no spin (and hence no magnetic moment), while odd-odd nuclei are often unstable (of course with many exceptions) and tend not to be used in microwave clocks. We begin by considering the contribution of the valence nucleon alone.

\subsection*{Model A}

Assuming all other nucleons in the nucleus are paired and ignoring polarization of the core (which will be considered in model B), the spin $I$, and hence magnetic moment of the nucleus, is entirely due to the total angular momentum of that external nucleon: $I = j = l + s$. One obtains the standard Schmidt formula for the magnetic moment:
\begin{align}
\label{eq:mu_valence}
\mu^0 &= g_{s} \left< s_z \right>^0 + g_l \left< l_{z} \right>^0 \\
\label{eq:Sz_schmidt}
\left< s_z \right>^0 &=
 \left\{ \begin{array}{ll}
  \frac{1}{2},    & \textrm{if}\quad j = l+\frac{1}{2} \vspace*{0.5em}\\
  -\frac{j}{2(j+1)}, & \textrm{if}\quad j = l-\frac{1}{2}
 \end{array} \right. \\
\left< l_z \right>^0 &=
 \left\{ \begin{array}{ll}
  j - \frac{1}{2},  & \textrm{if}\quad j = l+\frac{1}{2} \vspace*{0.5em}\\
  \frac{j(2j+3)}{2(j+1)}, & \textrm{if}\quad j = l-\frac{1}{2}
 \end{array} \right. \nonumber
\end{align}
The gyromagnetic factors are $g_l = 1$, $g_s = g_p = 5.586$ for a valence proton and $g_l = 0$, $g_s = g_n = -3.826$ for a valence neutron. It is these Schmidt values for \Sp\ and \Sn\ that are presented as the model A results in Tables~\ref{tab:spin_proton} and~\ref{tab:spin_neutron}.

Because light-quark masses are small, the $g$-factors are sensitive to $m_q/\LQCD$ mainly via $\pi$-meson loop corrections to the nuclear magnetic moments, which are proportional to $\pi$-meson mass $m_\pi \sim \sqrt{m_q\LQCD}$~\cite{flambaum04prd}. Because $m_\pi = 140\,\MeV$, the quark-mass contribution can be significant even in the chiral limit $m_q = 0$. Full details are given in~\cite{flambaum04prd} (see also~\cite{cloet08fbs}); the final results are
\begin{align}
\label{eq:deltag}
\frac{\delta g_p}{g_p} &= -0.087 \frac{\delta m_q}{m_q} -0.013 \frac{\delta m_s}{m_s} \\
\frac{\delta g_n}{g_n} &= -0.118 \frac{\delta m_q}{m_q} +0.0013 \frac{\delta m_s}{m_s} \nonumber
\end{align}
where variations in $m_q$ and $m_s$ (strange quark mass) are considered separately. Note that the $g$-factors are far more sensitive to $m_q$ than $m_s$. If one assumes the quark-masses as proportional to the vacuum expectation value of the Higgs field, then $\delta m_q/m_q$ should be approximately equal to $\delta m_s/m_s$. The total variation $\delta g/g$ would then be the sum of the variations in $m_q$ and $m_s$.

In is convenient at this stage to define parameters $K_p$ and $K_n$ by
\begin{equation}
\label{eq:K_def}
\frac{\delta \mu}{\mu} = \frac{\delta g_p}{g_p} K_p + \frac{\delta g_n}{g_n} K_n \,,
\end{equation}
where, according to~\eref{eq:mu_general},
\begin{equation}
K_p = \frac{g_p\Sp}{\mu} \quad \textrm{and} \quad
K_n = \frac{g_n\Sn}{\mu} \,.
\end{equation}
For example, for a valence proton, model A gives $K_p = g_p \Sp^0/\mu^0$ and $K_n = 0$. \Eref{eq:K_def} shows the connection between the spin expectation value and sensitivity to variation of fundamental constants. 

Combining Eqs.~\eref{eq:deltag} and \eref{eq:K_def} one may obtain the sensitivity of the magnetic moment $\mu$ to variation in light and strange quark masses for any nucleus with a single valence nucleon (more details may be found in \cite{flambaum06prc}). The result is characterized by the sensitivity coefficients $\kappa_q$ and $\kappa_s$, defined by
\begin{equation}
\label{eq:kappa_def}
\frac{\delta\mu}{\mu} = \kappa_q \frac{\delta m_q}{m_q} + \kappa_s \frac{\delta m_s}{m_s}.
\end{equation}
where
\begin{align*}
\kappa_q &= -0.118\, K_n - 0.087\, K_p \\
\kappa_s &= 0.0013\, K_n - 0.013\, K_p \,.
\end{align*}
In this paper we are interested in finding large differences in $\kappa=\kappa_q+\kappa_s$ between different isotopes of the same nucleus. With this in mind, we present in \Tref{tab:kappa} the $\kappa$ obtained using model A, via \Eref{eq:Sz_schmidt}, for all isotopes of atoms commonly used in microwave clocks. We defer discussion of the results to \Sec{sec:discussion}.

\subsection*{Model B}

\begin{table}[tb]
\caption{\label{tab:spin_proton} Average values of \Sp\ and \Sn\ for heavy nuclei with a valence proton. A: Schmidt values (no neutron contribution); B\,I: our preferred fit of experimental magnetic moment assuming separate conservation of $j_p$, $j_n$, and total $l$ and $s$; B\,II: fit $\mu$ assuming minimal transfer of spin from protons to neutrons. }
\begin{ruledtabular}
\begin{tabular}{cddddd}
 & \multicolumn{3}{c}{\Sp} & \multicolumn{2}{c}{\Sn} \\
\cline{2-4}
\cline{5-6}
Isotope & \multicolumn{1}{c}{A} & \multicolumn{1}{c}{B\,I} & \multicolumn{1}{c}{B\,II}
 & \multicolumn{1}{c}{B\,I} & \multicolumn{1}{c}{B\,II} \\
\hline
$^{133}$Cs &    
-0.389 & -0.286 & -0.297 & -0.103 & -0.092
\\    
$^{135}$Cs &    
-0.389 & -0.268 & -0.281 & -0.121 & -0.108
\\    
$^{137}$Cs &    
-0.389 & -0.255 & -0.269 & -0.134 & -0.120
\\    
$^{131}$Cs &    
0.500 & 0.351 & 0.367 & 0.149 & 0.133
\\    
$^{129}$Cs &    
0.500 & 0.345 & 0.362 & 0.155 & 0.138
\\    
$^{127}$Cs &    
0.500 & 0.341 & 0.358 & 0.159 & 0.142
\\    
$^{125}$Cs &    
0.500 & 0.335 & 0.353 & 0.165 & 0.147
\\    
$^{139}$Cs &    
-0.389 & -0.272 & -0.285 & -0.116 & -0.104
\\    
$^{123}$Cs &    
0.500 & 0.332 & 0.35 & 0.168 & 0.150
\\
\hline
$^{139}$La &
-0.389 & -0.262 & -0.276 & -0.127 & -0.113
\\
$^{137}$La &
-0.389 & -0.273 & -0.285 & -0.116 & -0.104
\\
\hline    
$^{223}$Fr &    
0.500 & 0.188 & 0.221 & 0.312 & 0.279
\\    
$^{221}$Fr &    
-0.357 & -0.272 & -0.281 & -0.085 & -0.076
\\    
$^{225}$Fr &    
0.500 & 0.176 & 0.211 & 0.324 & 0.289
\\    
$^{227}$Fr &    
0.500 & 0.346 & 0.363 & 0.154 & 0.137
\\
\hline    
$^{203}$Tl &    
0.500 & 0.361 & 0.376 & 0.139 & 0.124
\\    
$^{205}$Tl &    
0.500 & 0.363 & 0.377 & 0.137 & 0.123
\\    
$^{201}$Tl &    
0.500 & 0.359 & 0.374 & 0.141 & 0.126
\\    
$^{199}$Tl &    
0.500 & 0.358 & 0.373 & 0.142 & 0.127
\\    
$^{197}$Tl &    
0.500 & 0.356 & 0.371 & 0.144 & 0.129
\\    
$^{195}$Tl &    
0.500 & 0.356 & 0.371 & 0.144 & 0.129
\\    
$^{193}$Tl &    
0.500 & 0.357 & 0.372 & 0.143 & 0.128
\\    
$^{207}$Tl &    
0.500 & 0.391 & 0.403 & 0.109 & 0.097 
\\   
\end{tabular}
\end{ruledtabular}
\end{table}


\begin{table}[tb]
\caption{\label{tab:spin_neutron} Average values of \Sp\ and \Sn\ for nuclei with a valence neutron. A: Schmidt values (no proton contribution); B\,I: our preferred fit of experimental magnetic moment assuming separate conservation of $j_p$, $j_n$, and total $l$ and $s$; B\,II: fit $\mu$ assuming minimal transfer of spin from protons to neutrons. }
\begin{ruledtabular}
\begin{tabular}{cddddd}
 & \multicolumn{3}{c}{\Sn} & \multicolumn{2}{c}{\Sp} \\
\cline{2-4}
\cline{5-6}
Isotope
 & \multicolumn{1}{c}{A} & \multicolumn{1}{c}{B\,I} & \multicolumn{1}{c}{B\,II}
 & \multicolumn{1}{c}{B\,I} & \multicolumn{1}{c}{B\,II} \\
\hline
$^{129}$Xe &    
0.500 & 0.365 & 0.379 & 0.135 & 0.121
\\    
$^{131}$Xe &    
-0.300 & -0.246 & -0.252 & -0.054 & -0.048
\\    
$^{127}$Xe&    
0.500 & 0.332 & 0.35 & 0.168 & 0.150
\\ 
$^{133}$Xe&    
-0.300 & -0.26 & -0.264 & -0.04 & -0.036
\\ 
$^{135}$Xe&    
-0.300 & -0.271 & -0.274 & -0.029 & -0.026
\\        
\hline    
$^{171}$Yb &    
-0.167 & -0.150 & -0.151 & -0.017 & -0.015
\\    
$^{173}$Yb &    
-0.357 & -0.118 & -0.143 & -0.239 & -0.214
\\    
$^{169}$Yb &    
-0.389 & -0.137 & -0.163 & -0.252 & -0.226
\\    
$^{175}$Yb &    
0.500 & 0.181 & 0.215 & 0.319 & 0.285
\\    
$^{167}$Yb &    
-0.357 & -0.269 & -0.278 & -0.088 & -0.079
\\    
$^{163}$Yb &    
0.500 & 0.317 & 0.336 & 0.183 & 0.164
\\    
$^{165}$Yb &    
-0.357 & -0.252 & -0.263 & -0.106 & -0.094
\\    
$^{161}$Yb &    
0.500 & 0.311 & 0.331 & 0.189 & 0.169
\\    
\hline    
$^{199}$Hg &    
-0.167 & -0.151 & -0.153 & -0.016 & -0.014
\\    
$^{201}$Hg &    
0.500 & 0.339 & 0.356 & 0.161 & 0.144
\\    
$^{203}$Hg &    
-0.357 & -0.296 & -0.302 & -0.062 & -0.055
\\    
$^{197}$Hg &    
-0.167 & -0.154 & -0.155 & -0.013 & -0.012
\\    
$^{195}$Hg &    
-0.167 & -0.155 & -0.156 & -0.011 & -0.010
\\    
$^{193}$Hg &    
0.500 & 0.347 & 0.363 & 0.153 & 0.137
\\    
$^{191}$Hg &    
0.500 & 0.346 & 0.362 & 0.154 & 0.138
\\    
$^{189}$Hg &    
0.500 & 0.345 & 0.361 & 0.155 & 0.139
\\    
$^{205}$Hg &    
-0.167 & -0.162 & -0.163 & -0.004 & -0.004
\\    
\hline    
$^{225}$Ra &    
0.500 & 0.360 & 0.375 & 0.140 & 0.125
\\    
$^{223}$Ra &    
-0.300 & -0.196 & -0.207 & -0.104 & -0.093
\\    
$^{227}$Ra &    
-0.300 & -0.116 & -0.135 & -0.184 & -0.165
\\    
$^{229}$Ra &    
0.500 & 0.213 & 0.243 & 0.287 & 0.257
\\    
\hline    
$^{211}$ Rn &    
-0.167 & -0.162 & -0.163 & -0.004 & -0.004
\\    
$^{209}$ Rn &    
-0.357 & -0.294 & -0.301 & -0.063 & -0.056
\\    
$^{207}$ Rn &    
-0.357 & -0.292 & -0.299 & -0.065 & -0.058
\\    
$^{205}$ Rn &    
-0.357 & -0.290 & -0.297 & -0.067 & -0.060
\\    
\end{tabular}
\end{ruledtabular}
\end{table}

The shell model of the previous section is known to overestimate the magnetic moment determined from experiment. This can be understood as polarisation of the non-valence nucleons which acts to reduce the magnetic moment. Using the experimental value of the magnetic moment we can estimate this core polarisation, and thus improve our estimates for \Sp\ and \Sn. 

There are many ways to enact a reduction in magnetic moment from the Schmidt value $\mu^0$ to the experimental value $\mu$. The most efficient means is to assume that the spin-spin interaction transfers spin from the valence proton (neutron) to core neutrons (protons):
\begin{equation}
\label{eq:minimal_transfer}
(\Sp - \Sp^0) = -(\Sn - \Sn^0) = \frac{\mu - \mu^0}{g_p - g_n}
\end{equation}
where $\Sn^0$ and $\Sp^0$ are the Schmidt values from model A (one will necessarily be zero). The denominator $g_p - g_n = 9.412$ is a large number, so the required change in \Sp\ and \Sn\ to obtain the experimental $\mu$ is minimal.
The transferred spin goes from contributing with a gyromagnetic factor of $g_p=+5.586$ to $g_n=-3.826$. 

Of course nature may proceed in an entirely different fashion, for example by transferring spin of a valence proton to its orbital angular momentum, or in a rather more unlikely fashion by transferring valence proton spin to neutron orbital momentum. Even under extreme scenarios, it was shown in~\cite{flambaum06prc} that the difference in the correction to the $m_q$-dependence varied by less than 10\%. This stability of the final result to the assumptions enables us to ignore the detailed nuclear forces and use ``heuristic'' fitting of $\mu$ to obtain values of \Sp\ and \Sn.

The preferred method of fitting $\mu$ in~\cite{flambaum06prc} is to assume that the total $z$-projection of proton and neutron angular momenta, $j_{pz}$ and $j_{nz}$, are separately conserved, and that total spin and orbital angular momenta $z$-projections, $\Sp + \Sn$ and $\left<l_p\right> + \left<l_n\right>$, are also separately conserved (corresponding to neglect of the spin-orbit interaction). The final result gives values somewhere in between those of the ``extreme'' assumptions. Then
\begin{align*}
 \left<s_z\right>^0 &= \Sp + \Sn \\
 \left<j_{pz}\right>  &= \left< l_p \right> + \Sp
\end{align*}
where $\left<j_{pz}\right> = I$ for valence proton and is zero for a valence neutron and $\left<s_z\right>^0$ is the Schmidt value for the spin of the valence nucleon~\eref{eq:Sz_schmidt}. Manipulation of these expressions with \eref{eq:mu_general} gives
\begin{align}
\label{eq:Sz_modelB}
\Sn &= \frac{\mu - \left<j_{pz}\right> - (g_p-1)\left<s_z\right>^0}{g_n - g_p + 1} \\
\Sp &= \left<s_z\right>^0 - \Sn \,. \nonumber
\end{align}
These neutron and proton spins are presented for isotopes of elements used (or proposed) in microwave clocks as the BI column of Tables~\ref{tab:spin_proton} and~\ref{tab:spin_neutron}. The ``minimal change'' model~\eref{eq:minimal_transfer} is presented as the BII column, and serves to indicate the expected uncertainty in our calculation.
Sensitivity coefficients $\kappa$ calculated using values of \Sp\ and \Sn\ given by \eref{eq:Sz_modelB} (corresponding to model BI) are given in the model B column of \Tref{tab:kappa}.

\subsection*{Model C}

To calculate the expectation values \Sp\ and \Sn, model B is the best we can do without venturing into highly model-dependent and complex nuclear calculations. However, we should also estimate how these spin expectation values will change under a variation of quark mass, and include this in our calculation of $\kappa_q$. Under the assumption of conservation of nuclear spin (model B), we may define a parameter $b$ in the case of a valence proton by
\begin{align}
\label{eq:b_def}
\Sn &= b \left< s_z \right>^0 \\
\Sp &= (1-b) \left< s_z \right>^0 \,.\nonumber
\end{align}
The coefficient $b$ is determined by the spin-spin interaction and is estimated in model B by~\eref{eq:Sz_modelB}. In the case of a valence neutron, we should define the small parameter $b$ by $\Sn = (1-b) \left< s_z \right>^0$ instead of \eref{eq:b_def}.

Under a variation of $m_q$, the parameter $b$ will change according to
\begin{equation}
\label{eq:b_dependence}
\frac{\delta b}{b} = -0.11\frac{\delta m_q}{m_q} \,.
\end{equation}
Full details are presented in~\cite{flambaum06prc}. Briefly, the spin-orbit splitting remains finite in the chiral limit, and so it only has a weak dependence on $m_q/\LQCD$. The spin-spin interaction is calculated in perturbation theory (to leading order) under the assumption that the major dependence comes from $\pi$-meson exchange (1/3 of the spin-spin interaction) and $\rho$-meson exchange (the remaining 2/3). If one instead assumes that the spin-spin interaction is dominated entirely by $\pi$-meson exchange, the coefficient in~\eref{eq:b_dependence}, $-0.17$, is not too different.

We include the effect of variation of spin expectation values on magnetic moment by generalising our previous formula~\eref{eq:K_def} to obtain
\begin{equation}
\label{eq:Kb_def}
\frac{\delta \mu}{\mu} = \frac{\delta g_n}{g_n}K_n + \frac{\delta g_p}{g_p}K_p + \frac{\delta b}{b}K_{b}
\end{equation}
For a valence proton:
\begin{equation}
\label{eq:Kb_proton}
K_{b_p} = \frac{(g_n - g_p +1)\Sn}{\mu} \,,
\end{equation}
while for a valence neutron:
\begin{equation}
\label{eq:Kb_neutron}
K_{b_n} = \frac{(g_p - g_n - 1)\Sp}{\mu} \,.
\end{equation}
The final sensitivity of our magnetic moment on $m_q$ is given in model C by
\begin{equation}
\label{eq:kappa_q}
\kappa = -0.117\, K_n - 0.100\, K_p - 0.11\, K_b \,.
\end{equation}
Calculations of $\kappa$ according to model C are presented in \Tref{tab:kappa} along with results of the other two methods.


\begin{table*}[htb]
\caption{\label{tab:kappa} Summary of magnetic moment sensitivity to quark-mass variation, $\kappa=\kappa_q+\kappa_s$. Columns A, B and C refer to the different methods as described in Section II: model C is the best value, while the difference between B and C gives an indication of the theoretical uncertainty. The difference in $\kappa$ values for different isotopes of a particular element (relative to the first isotope) are given for models B and C. The isotope entries are given in order of half-life length, beginning with stable isotopes.}
\begin{ruledtabular}
\begin{tabular}[b]{crdrddddd}
Atom & \multicolumn{1}{c}{J$^\pi$} & \multicolumn{1}{c}{$\mu$~\cite{stone05adndt}} & \multicolumn{1}{c}{Half-life} & \multicolumn{1}{c}{A} & \multicolumn{1}{c}{B} & \multicolumn{1}{c}{C} & \multicolumn{1}{c}{$\delta$B} & \multicolumn{1}{c}{$\delta$C} \\
\hline\hline
 $^{27}$Al & $5/2^+$ & +3.642 & Stable & -0.058 & -0.039 & -0.004 & - & - \\
 $^{25}$Al\footnotemark[1] & $5/2^+$ & +3.646 & 7.18 s & -0.058 & -0.039 & -0.004 & 0 & 0 \\
 \hline
 $^{69}$Ga & $3/2^-$ & +2.017 & Stable & -0.074 & -0.033 & 0.064 & - & - \\     
 $^{71}$Ga & $3/2^-$ & +2.562 & Stable & -0.074 & -0.052 & 0.001 & -0.019 & -0.063 \\     
 $^{67}$Ga & $3/2^-$ & +1.851 & 78.3 h & -0.074 & -0.026 & 0.090 & 0.007 & 0.026 \\     
 \hline     
 $^{85}$Rb & $5/2^-$ & +1.353 & Stable & 0.231 & 0.104 & 0.064 & - & - \\     
 $^{87}$Rb & $3/2^-$ & +2.751 & $4.9 \times 10^{10}$ y & -0.074 & -0.056 & -0.015 & -0.160 & -0.079 \\     
 $^{83}$Rb & $5/2^-$ & +1.425 & 86.2 d & 0.231 & 0.093 & 0.049 & -0.011 & -0.015 \\     
 $^{81}$Rb & $3/2^-$ & +2.060 & 4.58 h & -0.074 & -0.035 & 0.058 & -0.139 & -0.006 \\     
 $^{79}$Rb & $5/2^+$ & +3.358 & 23 m & -0.058 & -0.032 & 0.015 & -0.136 & -0.049 \\     
 $^{77}$Rb & $3/2^-$ & +0.654 & 3.8 m & -0.074 & 0.146 & 0.674 & 0.042 & 0.61 \\     
 \hline     
 $^{111}$Cd & $1/2^+$ & -0.595 & Stable & -0.117 & -0.111 & 0.133 & - & - \\     
 $^{113}$Cd & $1/2^+$ & -0.622 & $9 \times 10^{15}$ y & -0.117 & -0.111 & 0.117 & 0 & -0.016 \\     
 $^{109}$Cd & $5/2^+$ & -0.828 & 453 d & -0.117 & -0.113 & 0.031 & -0.002 & -0.102 \\     
 $^{115}$Cd & $1/2^+$ & -0.648 & 53.4 h & -0.117 & -0.111 & 0.103 & 0 & -0.03 \\     
 $^{107}$Cd & $5/2^+$ & -0.615 & 6.5 h & -0.117 & -0.111 & 0.121 & 0 & -0.012 \\     
 $^{105}$Cd & $5/2^+$ & -0.739 & 56 m & -0.117 & -0.112 & 0.062 & -0.001 & -0.071 \\     
 $^{103}$Cd & $5/2^+$ & -0.81(3) & 7.3 m & -0.117 & -0.113 & 0.037 & -0.002 & -0.096 \\     
 \hline     
 $^{133}$Cs & $7/2^+$ & +2.582 & Stable & 0.127 & 0.044 & 0.007 & - & - \\     
 $^{135}$Cs & $7/2^+$ & +2.732 & $3 \times 10^{6}$ y & 0.127 & 0.035 & -0.006 & -0.009 & -0.013 \\     
 $^{137}$Cs & $7/2^+$ & +2.851 & 30.17 y & 0.127 & 0.029 & -0.014 & -0.015 & -0.022 \\     
 $^{131}$Cs & $5/2^+$ & +3.53(2) & 9.69 d & -0.058 & -0.037 & 0.002 & -0.081 & -0.005 \\     
 $^{129}$Cs & $1/2^+$ &+1.491 & 32.3 h & -0.100 & -0.083 & 0.013 & -0.127 & 0.006 \\     
 $^{127}$Cs & $1/2^+$ & +1.459 & 6.2 h & -0.100 & -0.082 & 0.018 & -0.126 & 0.011 \\     
 $^{125}$Cs & $1/2^+$ & +1.409 & 45 m & -0.100 & -0.081 & 0.027 & -0.125 & 0.020 \\     
 $^{139}$Cs & $7/2^+$ & +2.696 & 9.4 m & 0.127 & 0.037 & -0.003 & -0.007 & -0.010 \\     
 $^{123}$Cs & $1/2^+$ & +1.377 & 5.8 m & -0.100 & -0.080 & 0.033 & -0.124 & 0.026 \\
 \hline
 $^{139}$La & $7/2^+$ & +2.783 & Stable & 0.127 & 0.032 & -0.010 & - & - \\
 $^{137}$La & $7/2^+$ & +2.695 & $6 \times 10^4$ y & 0.127 & 0.037 & -0.003 & 0.005 & 0.007 \\
 \hline     
 $^{171}$Yb & $1/2^-$ & +0.494 & Stable & -0.117 & -0.116 & -0.084 & - & - \\     
 $^{173}$Yb & $5/2^-$ & -0.648 & Stable & -0.117 & -0.125 & -0.467 & -0.009 & -0.383 \\     
 $^{169}$Yb & $7/2^+$ & -0.635 & 32.0 d & -0.117 & -0.126 & -0.494 & -0.010 & -0.410 \\     
 $^{175}$Yb\footnotemark[1] & $7/2^-$ & +0.768 & 4.18 d & -0.117 & -0.126 & -0.510 & -0.010 & -0.426 \\     
 $^{167}$Yb & $5/2^-$ & +0.623 &17.5 m & -0.117 & -0.113 & 0.018 & 0.003 & 0.102 \\     
 $^{165}$Yb & $5/2^-$ & +0.478 & 9.9 m & -0.117 & -0.112 & 0.093 & 0.004 & 0.177 \\     
 $^{161}$Yb & $3/2^-$ & -0.327 & 4.2 m & -0.117 & -0.103 & 0.430 & 0.013 & 0.514 \\     
 \hline     
 $^{199}$Hg & $1/2^-$ & +0.506 & Stable & -0.117 & -0.116 & -0.087 & - & - \\     
 $^{201}$Hg & $3/2^-$ & -0.560 & Stable & -0.117 & -0.110 & 0.156 & 0.006 & 0.243 \\     
 $^{203}$Hg & $5/2^-$ & +0.849 & 46.8 d & -0.117 & -0.115 & -0.048 & 0.001 & 0.039 \\     
 $^{197}$Hg & $1/2^-$ & +0.527 & 64.1 h & -0.117 & -0.116 & -0.093 & 0 & -0.006 \\      
 $^{195}$Hg & $1/2^-$ & +0.541 & 9.9 h & -0.117 & -0.116 & -0.097 & 0 & -0.009  \\      
 $^{193}$Hg & $3/2^-$ & -0.628 & 3.80 h & -0.117 & -0.111 & 0.114 & 0.005 & 0.202  \\      
 $^{191}$Hg & $3/2^-$ & -0.618 & 49 m & -0.117 & -0.111 & 0.120 & 0.005 & 0.207  \\      
 $^{189}$Hg & $3/2^-$ & -0.609 & 7.6 m & -0.117 & -0.111 & 0.125 & 0.005 & 0.212  \\      
\end{tabular}
\footnotetext[1]{The sign of the magnetic moment is unknown: we have assumed positive sign.}
\end{ruledtabular}
\end{table*}


\section{Hyperfine transitions}


To date, the most accurate probe of variation of fundamental constants in the laboratory comes from atomic clocks. Microwave clocks based on the hyperfine interaction, such as the caesium fountain from which the SI time unit is defined, have achieved precision at parts in $10^{15}$ at several laboratories worldwide. To find the sensitivity of such clocks to potential variation of fundamental constants, we start with the Hamiltonian of the hyperfine transition which can be expressed as
\begin{equation}
H_\textsl{hfs} = A\, \vect{I} \cdot \vect{J}
\end{equation}
where $\vect{I}$ and $\vect{J}$ are the nuclear and electron angular momenta, respectively. The $A$ coefficient can be expressed as
\begin{equation}
A = \textrm{const} \left( \frac{me^4}{\hbar^2} \right)
\left[\alpha^2 F_\textrm{rel}(Z\alpha) \left(1 - \delta_h \right) \right] \mu \frac{m_e}{m_p}
\end{equation}
where $\delta_h(r_n, Z\alpha)$ is the small effect of finite nuclear radius $r_n$~\cite{dinh09pra} and $F_\textrm{rel}(Z\alpha)$ is the effect of relativistic corrections to the electron wavefunction at the nucleus. When comparing hyperfine transitions of two different nuclei, sensitivity to variation of fundamental constants will arise due to differences in $\mu$, $F_\textrm{rel}$, and $\delta_h$.

$F_\textrm{rel}$ and $\delta_h$ are sensitive to details of the electron wavefunction, but are almost identical for different isotopes of the same element. Nevertheless, when comparing hyperfine transitions across different elements $F_\textrm{rel}$ and $\delta_h$ may be included along with $\mu$. $\delta_h$ and $\mu$ are mostly sensitive to quark mass, while $F_\textrm{rel}$ is sensitive to variation in the value of $\alpha$. Therefore the effect of variation of these fundamental constants on a hyperfine transition frequency $\omega$ (when measured against another hyperfine frequency) can be expressed
\begin{equation}
\label{eq:hyperfineeffects}
\frac{\delta\omega}{\omega} = K_\textrm{rel} \frac{\delta\alpha}{\alpha} + (\kappa_q + k_{hq}) \frac{\delta m_q}{m_q} \,.
\end{equation}
Here $K_\textrm{rel}$ and $k_{hq}$ represent sensitivities of $\omega$ to $F_\textrm{rel}$ and $\delta_h$, respectively.

For hyperfine transitions of $s$-wave electrons the following formulae may be used:
\begin{equation}
\label{eq:K_rel_analytical}
K_\textrm{rel} \approx \frac{(Z\alpha)^2(12\gamma^2 - 1)}{\gamma^2(4\gamma^2-1)}
\end{equation}
\begin{equation}
\label{eq:k_hq}
k_{hq} = -0.3 \times \frac{(2\gamma - 1)\delta_h}{\delta_h}
\end{equation}
where $\gamma = \sqrt{1 - (Z\alpha)^2}$ and
\begin{equation}
\label{eq:delta_h_analytical}
\delta_h \approx 1.995 \left({Z r_n}/{a_B}\right)^{2\gamma-1}\,.
\end{equation}
Comparison with numerical calculations shows that \Eref{eq:K_rel_analytical} tends to underestimate the $\alpha$-dependence by a small amount (less than $\sim15\%$), while \eref{eq:delta_h_analytical} includes a fitting parameter to the calculations that makes it accurate to within a few percent for all tested ions. In \Tref{tab:khq} we present the results of previous numerical calculations of $K_\textrm{rel}$~\cite{flambaum06prc} and $k_{hq}$~\cite{dinh09pra}. In the case where the microwave clock is based on hyperfine splitting of a $p$-wave electron, such as proposed Al and Ga clocks, both corrections may be neglected.

\begin{table}[tb]
\caption{Sensitivity to quark mass variation due to nuclear size contribution, $k_{hq}$, and sensitivity to $\alpha$ variation, $K_\textrm{rel}$, for atoms and ions with $s$-wave valence electrons (these are defined by \Eref{eq:hyperfineeffects}). \label{tab:khq}}
\begin{tabular}{ldd}
\hline\hline
Atom & \multicolumn{1}{c}{$k_{hq}$} & \multicolumn{1}{r}{$K_\textrm{rel}$} \\
\hline
Rb     & -0.003 & 0.34 \\
Cd$^+$ & -0.005 & 0.6 \\
Cs     & -0.007 & 0.83 \\
Yb$^+$ & -0.014 & 1.5 \\
Hg$^+$ & -0.023 & 2.28 \\
\hline\hline
\end{tabular}
\end{table}

\section{Discussion}
\label{sec:discussion}

Using \Tref{tab:kappa} we may identify good isotopes for microwave clock studies of variation of fundamental constants. The largest difference within a single element, $\delta C$ in the table, is seen in ytterbium. Comparison of clocks based on hyperfine transitions in $^{161}$Yb and $^{169}$Yb should yield a relative sensitivity of $\delta\kappa=0.924$. However one should be cautious here. Because the valence nucleon is a neutron, the lowest order result (model A) is zero. The entire effect comes from the contribution of polarisation of the core protons (model B) and changes of this polarisation due to quark mass variation (model C). Thus, despite the apparent stability of \eref{eq:b_dependence} as described earlier, the strong change in sensitivity $\kappa$ between models B and C should be treated with caution.

The same argument suggests caution in interpretation of the $^{201}$Hg -- $^{199}$Hg comparison~\cite{burt09fsm}. The sensitivity factor of $\kappa_q=0.24$ predicted by model C represents a large departure from the model B value of 0.006. A more satisfactory result from this perspective is obtained from the $^{87}$Rb -- $^{85}$Rb clock comparison, where models B and C give $\kappa=-0.16$ and $-0.08$, respectively. The variation between the models is not so strong, yet the final sensitivity factor, $\kappa = -0.08$, is still reasonably large.

We have also included a number of comparisons between isotopes of different atoms in \Tref{tab:crossatom}. $\delta\kappa$ defines the sensitivity of a ratio, $\omega_1/\omega_2$, of hyperfine transition frequencies of two isotopes to variation of quark mass:
\begin{equation}
\label{eq:deltakappa}
\frac{\delta(\omega_1/\omega_2)}{(\omega_1/\omega_2)} = (\kappa_1-\kappa_2) \frac{\delta m_q}{m_q} = \delta\kappa \frac{\delta m_q}{m_q} \,.
\end{equation}
These comparisons exhibit strong sensitivity as well as stability across the models. Once again, model C represents our most complete calculation, while the difference between B and C gives an indication of the theoretical uncertainty. The final result includes the contribution of nuclear radius, $\delta\kappa = \delta C + \delta k_{hq}$.

\begin{table}[!htb]
\caption{\label{tab:crossatom} Comparisons of sensitivity to quark-mass variation, $\kappa$, for isotopes of different atoms (see \Eref{eq:deltakappa}). Our final result is $\delta\kappa = \delta\textrm{C} + \delta k_{hq}$.}
\begin{ruledtabular}
\begin{tabular}{ccccc}
Comparison: & \multicolumn{1}{c}{$\delta$B} & \multicolumn{1}{c}{$\delta$C} & \multicolumn{1}{c}{$\delta k_{hq}$} & \multicolumn{1}{c}{$\delta\kappa$} \\
\hline
$^{133}$Cs -- $^{199}$Hg & 0.160 & 0.095 & 0.016 & 0.111
\\
$^{85}$Rb -- $^{199}$Hg  & 0.220 & 0.151 & 0.020 & 0.171
\\
$^{133}$Cs -- $^{173}$Yb & 0.169 & 0.474 & 0.007 & 0.481
\\
$^{133}$Cs -- $^{169}$Yb & 0.170 & 0.501 & 0.007 & 0.508
\\
$^{85}$Rb -- $^{173}$Yb  & 0.229 & 0.531 & 0.011 & 0.542
\\
\end{tabular}
\end{ruledtabular}
\end{table}


\section*{Acknowledgements}
We thank Rong Wei and Eric Burt for useful discussions and for motivating this work.
This work was supported by the Australian Research Council.

\bibliography{references}

\end{document}

%% file: phys_jdf.tex
\def\adndt{At. Data Nucl. Data Tables}

\def\epjc{Eur. Phys. J. C}

\def\epjst{Eur. Phys. J. Spec. Topics}

\def\fbs{Few-Body Sys.}

\def\ijmpa{Int. J. Mod. Phys. A}

\def\jcap{J. Cosmol. Astropart. Phys.}

\def\jpcs{J. Phys.: Conf. Ser.}

\def\mnras{Mon. Not. R. Astron. Soc.}

\def\npb{Nucl. Phys. B}

\def\plb{Phys. Lett. B}

\def\pra{Phys. Rev. A}

\def\prc{Phys. Rev. C}
\def\prd{Phys. Rev. D}

\def\prl{Phys. Rev. Lett.}
\def\prep{Phys. Rep.}